\documentclass[twocolumn]{aastex63}
\pdfoutput=1 
\usepackage{amsmath,amstext}
\usepackage{blindtext}
\usepackage[T1]{fontenc}
\usepackage{apjfonts} 
\usepackage[figure,figure*]{hypcap}

\newcommand{\I}{\protect\small I \normalsize $\!\!$}

\newcommand{\HI}{\mbox{H\,{\sc i}}}

\newcommand{\Htwo}{H$_{2}$}

\newcommand{\twCO}{$^{12}$CO}

\newcommand{\pcmcub}{\mbox{${\rm cm^{-3}}$}}

\newcommand{\kmps}{\mbox{${\rm km\;s^{-1}}$}}

\newcommand{\lsim}{\mbox{$\mathrel{\vcenter{\hbox{\ooalign{\raise3pt\hbox{$<$}\crcr \lower3pt\hbox{$\sim$}}}}}$}}
\newcommand{\gsim}{\mbox{$\mathrel{\vcenter{\hbox{\ooalign{\raise3pt\hbox{$>$}\crcr \lower3pt\hbox{$\sim$}}}}}$}}
\shorttitle{Physics of the ''CO-Dark'' Gas}
\shortauthors{Busch et al.}

\submitjournal{ApJ}
\accepted{August 9th, 2019}

\draft

\begin{document}

\title{The Structure of Dark Molecular Gas in the Galaxy - II. \\ Physical State of ''CO-Dark'' Gas in the Perseus Arm}

\author{Michael P. Busch}
\thanks{To whom correspondence should be addressed: mpbusch@jhu.edu \\
MPB is a National Science Foundation Graduate Fellow}
\affiliation{Department of Physics and Astronomy, Johns Hopkins University, 3400 North Charles Street, Baltimore, MD 21218, USA}
\author{Ronald J. Allen}
\affiliation{Space Telescope Science Institute, 3700 San Martin Drive, Baltimore, MD 21218, USA}
\affiliation{Department of Physics and Astronomy, Johns Hopkins University, 3400 North Charles Street, Baltimore, MD 21218, USA}
\author{Philip D. Engelke} 
\affiliation{Department of Physics and Astronomy, Johns Hopkins University, 3400 North Charles Street, Baltimore, MD 21218, USA}
\author{David E. Hogg}
\affiliation{National Radio Astronomy Observatory, 520 Edgemont Road, Charlottesville, VA 22903, USA}
\author{David A. Neufeld}
\affiliation{Department of Physics and Astronomy, Johns Hopkins University, 3400 North Charles Street, Baltimore, MD 21218, USA}
\author{Mark G. Wolfire}
\affiliation{Department of Astronomy, University of Maryland, College Park, MD 20742, USA}

\begin{abstract}
We report the results from a new, highly sensitive ($\Delta T_{mb} \sim 3 $mK) survey for thermal OH emission at 1665 and 1667 MHz over a dense, 9 x 9-pixel grid covering a $1\degr \times 1\degr $ patch of sky in the direction of $l = 105\fdg00, b = +2\fdg50$  towards the Perseus spiral arm of our Galaxy. We compare our Green Bank Telescope (GBT) 1667 MHz OH results with archival \twCO(1-0) observations from the Five College Radio Astronomy Observatory (FCRAO) Outer Galaxy Survey within the velocity range of the Perseus Arm at these galactic coordinates. Out of the 81 statistically-independent pointings in our survey area, 86\% show detectable OH emission at 1667 MHz, and 19\% of them show detectable CO emission. We explore the possible physical conditions of the observed features using a set of diffuse molecular cloud models. In the context of these models, both OH and CO disappear at current sensitivity limits below an A$_{\rm v}$ of 0.2, but the CO emission does not appear until the volume density exceeds 100-200 \pcmcub.  These results demonstrate that a combination of low column density A$_{\rm v}$ and low volume density $n_{H}$ can explain the lack of CO emission along sight lines exhibiting OH emission. The 18-cm OH main lines, with their low critical density of $n^{*}$ $ \sim 1 $ \pcmcub, are collisionally excited over a large fraction of the quiescent galactic environment and, for observations of sufficient sensitivity, provide an optically-thin radio tracer for diffuse H$_2$.
\end{abstract}
\keywords{Galaxy: disk --- ISM: molecules --- ISM: structure --- local interstellar matter --- radio lines: ISM --- surveys}

\section{Introduction}
The formation of molecular hydrogen gas \Htwo\ from atomic hydrogen \HI\ in galaxies is widely considered to be a critical step for the formation of new stars from the interstellar gas, and hence it is one of the most important processes that occurs in the interstellar medium (ISM). Unfortunately, as a symmetric molecule without a dipole moment, \Htwo\ is practically invisible in emission at the temperature range of 10-100K expected for the bulk of the ISM in the Galaxy, and indirect estimates are required that make use of surrogate tracers. The most universally-accepted surrogate tracer for \Htwo\ in the ISM is the lowest-energy rotational spectral line of \twCO(1-0) at $\lambda = 3 $mm. This line is relatively bright and easily observed, often with instrumentation designed specifically for that purpose. CO observations are commonly used jointly with an empirically-derived conversion factor, usually  called the ``X-factor'' \citep[see][for a review]{Bolatto2013TheFactor}. The strength of the 3-mm CO line emission is measured in units of K $\kmps$ and, by multiplying this line strength by the X-factor, one directly obtains an estimate of the \Htwo\ column density. 

A growing body of observational evidence points to an extra component of the ISM not traced by either the 21-cm \HI\ line or \twCO(1-0) emission. This excess component is usually referred to as ``dark gas'' \citep{Grenier2005UnveilingNeighborhood, Wolfire2010TheGas}. This dark gas may be a large fraction of the total molecular gas content in the Galaxy \citep{Pineda2013AComponents, Li2015QuantifyingGas}.  Quantifying how much dark gas exists is therefore of great interest in an effort to calibrate the X-factor and work towards a single prescription for scaling between a molecular tracer line emission and an accurate total molecular mass of \Htwo. \citet{Wolfire2010TheGas} constructed models of molecular cloud surfaces and determined that dark gas in molecular cloud surfaces can amount to $\sim 30\%$ of the total molecular mass of the cloud. Here, the column density is large enough so that \Htwo\ has sufficient column to remain self-shielded against the ambient UV flux, but CO is photodissociated and the carbon is the form of C or C$^{+}$. However, a large fraction of the faint CO gas might also arise in the diffuse ISM \citep{Papadopoulos2002MolecularDistances, Liszt2010TheGas, Allen2012Faint5circ, Allen2015The+1deg, Xu2016EvolutionTaurusb}.

Initially discovered in absorption at centimeter radio wavelengths in the diffuse ISM\footnote{Additional historical information on the discovery and early observations of the 18-cm radio lines of OH both in emission and absorption can be found in \cite{Allen2012Faint5circ} and \cite{Allen2015The+1deg}.},  OH has also recently been detected in the far-IR \citep{Wiesemeyer2016Far-infraredClouds}, also in absorption. The work reported here mainly concerns new 18-cm emission observations in the outer Galaxy. \cite{Allen2012Faint5circ} reported faint and widespread 18-cm OH emission in a blind survey of a small region in the second quadrant of the Galactic plane using the 25-m radio telescope at Onsala, Sweden. However, owing to spectrometer limitations and radio interference in the spectra, the Onsala blind survey was limited to the 1667 MHz OH line and to within 2 kpc of the Sun. Similar results were presented by \cite{Dawson2014SPLASH:Region} with the SPLASH survey using the Parkes Telescope; however, their survey of OH encountered high levels of background synchrotron continuum emission from the inner Galaxy at levels approaching the typical excitation temperatures of the 18-cm OH lines, effectively suppressing the extended OH emission outside of the `CO-bright' clouds. In the outer Galaxy, we can typically avoid this issue as the integrated background synchrotron emission is much weaker. More recent observations in the outer Galaxy by \citet[][hereafter Paper 1]{Allen2015The+1deg} using the GBT have shown main-line (1665 and 1667 MHz) OH emission to be present in regions largely devoid of CO emission, strongly suggesting that OH may be a good tracer of the dark gas. The main lines of OH emission in these regions are observed to be in the 5:9 ratio characteristic of optically-thin emission lines with level populations in local thermodynamic equilibrium (LTE). This makes calculating a column density of OH relatively straightforward as long as a good estimate for the excitation temperatures can be found. With an OH column density known from emission observations, the \Htwo\ column density can be directly estimated using the N(\Htwo)/N(OH) ratio of approximately $10^{-7}$ measured from UV absorption data \citep{Weselak2010TheMolecules,Nguyen2018Dust-GasISM, Engelke2018OHW5}.

In this paper we present the results of a densely-sampled, blind, highly-sensitive GBT emission survey of the two main OH lines in a one-square-degree field of the Perseus Arm. The purpose of this survey is to further explore the apparent connection of extended OH emission to the so-called 'dark molecular gas', and attempt to resolve structures of diffuse molecular gas on a large scale that are otherwise invisible to CO observations.

In Paper 1, we were able to compare sparse OH survey measurements taken with the GBT (FWHM $\sim$ 7.6') with spectra provided directly from archives of the CFA CO survey \citep{Dame2001TheSurvey}, since both data sets were observed at closely similar angular resolution ($\sim 7.6'$ for the GBT vs $\sim 8.4'$ for the CfA telescopes). However, the Five College Radio Astronomy Observatory (FCRAO) \twCO\ ``Outer Galaxy Survey'' \citep{Heyer1998TheGalaxy} has significantly higher angular resolution ($\sim 50\arcsec$) than the CFA \twCO\ survey, and hence significantly better sensitivity when smoothed to the GBT beam. We have therefore compared our OH data with a smoothed version of the FCRAO data. This allows for a direct, dense, observational comparison between these two important molecular tracers with comparable sensitivities for the first time.

It was observed in Paper 1 that OH emission was widespread, and that there was varying structure to the emission profiles on scales of $\sim$ 30 pc at the distance of the Perseus Arm feature. In an attempt to resolve the structure of this emission a new observing program was undertaken to increase the coverage of the original 2015 sparse survey. In this paper we therefore chose to restrict our analysis to the Perseus Arm feature in particular in an attempt to map the OH emission spatially and compare it to the CO emission at the $\sim$ 7 pc scale, as we show in Sect. \ref{morphology}. Due to the absence of the kinematic distance ambiguity in the outer Galaxy, we are able to differentiate components of the OH spectra (see Fig. \ref{fig:exampleSpectrum}). The distance of the Perseus Arm offered us the best consistently apparent option to measure this structure. The availability of accurate parallax distances in this direction (see Sect. \ref{distancetoarm}) made this spatial comparison possible. Statistical and spatial comparisons of OH and CO emission between the local and inter-arm features would also be interesting and will be the subject of future papers.

\section{Observations and Data}\label{Observations}

\begin{figure}[t]
\centering
\includegraphics[width=0.45\textwidth]{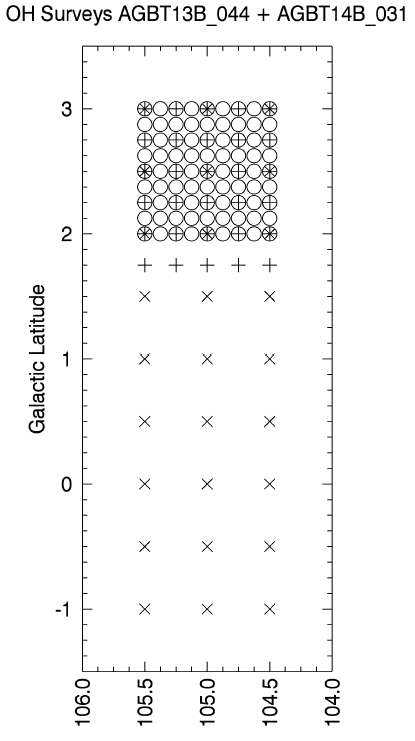} 
\caption{The blind survey areas discussed in this work. The ``X'' markers indicate the original 3x9 `sparse' survey carried out with the GBT ACS spectrometer in program AGBT13B\_044 as reported in Paper 1. The set of 9 ``X'' sightlines in the range of $+2\degr$ to $+3\degr$ were later re-observed using the newer VEGAS backend for consistency with the remaining sightlines. The "+" markers indicate the next 5x6 survey carried out in program AGBT14B\_031. As unexpected but significant L band GBT observing time became available, we changed the observing goal to sample an entire square degree. The "O" markers indicate the final "One-Square-Degree" 9x9 dense 81-point grid, also carried out under AGBT14B\_031. These data are the subject of this paper.}
\label{fig:osdArea}
\end{figure}

We carried out highly-sensitive observations of main-line OH emission at 18 cm with the Robert C. Byrd Green Bank Telescope (GBT) in project AGBT14B\_031 at the 81 locations indicated with small circles in Fig.\ref{fig:osdArea}. For comparison with the OH data, we used the \twCO(1-0) data from the FCRAO survey, as explained above.

\subsection{OH Observations} \label{data}

Paper 1 reported the results of a 'sparse' survey with the GBT at $0.5^{\circ}$ spacing interval at sightlines indicated with "X" in Fig.\ref{fig:osdArea}. The observations discussed in this paper were made between September 11, 2015 and January 31, 2016, also with the GBT. The receivers selected are situated at the Gregorian Focus and operate in the frequency range 1.15 to 1.73 GHz (L-Band). There is one beam on the sky, with dual polarizations. The amplifiers are cooled Field Effect Transistors (FET) with an effective system temperature of 20 K or less in good weather and for intermediate elevations of the GBT. These observations were made with linear polarization.

This new OSD survey covers a $9 \times 9$ grid of GBT sightlines, which roughly corresponds to $60 \times 60$ pc region at a 3.2 kpc distance to the Perseus Arm, centered on l = 105.0\degr, b = 2.5\degr\ and at intervals of 0.125\degr\ = $7.5'$, closely corresponding to the angular resolution of the GBT at the frequency of the OH main lines (FWHM $\sim 7.6'$). At each position 12 individual scans, each of duration 10 minutes, were made. For an individual scan, in one polarization, the effective integration time was approximately 292 seconds, and the expected rms in a single channel in the corresponding spectrum is approximately 34 mK. The effective integration time of the final spectrum after combination of the data from all scans, in both polarizations, is just under two hours.

\begin{figure*}[!ht]
\centering
\includegraphics[width=\textwidth]{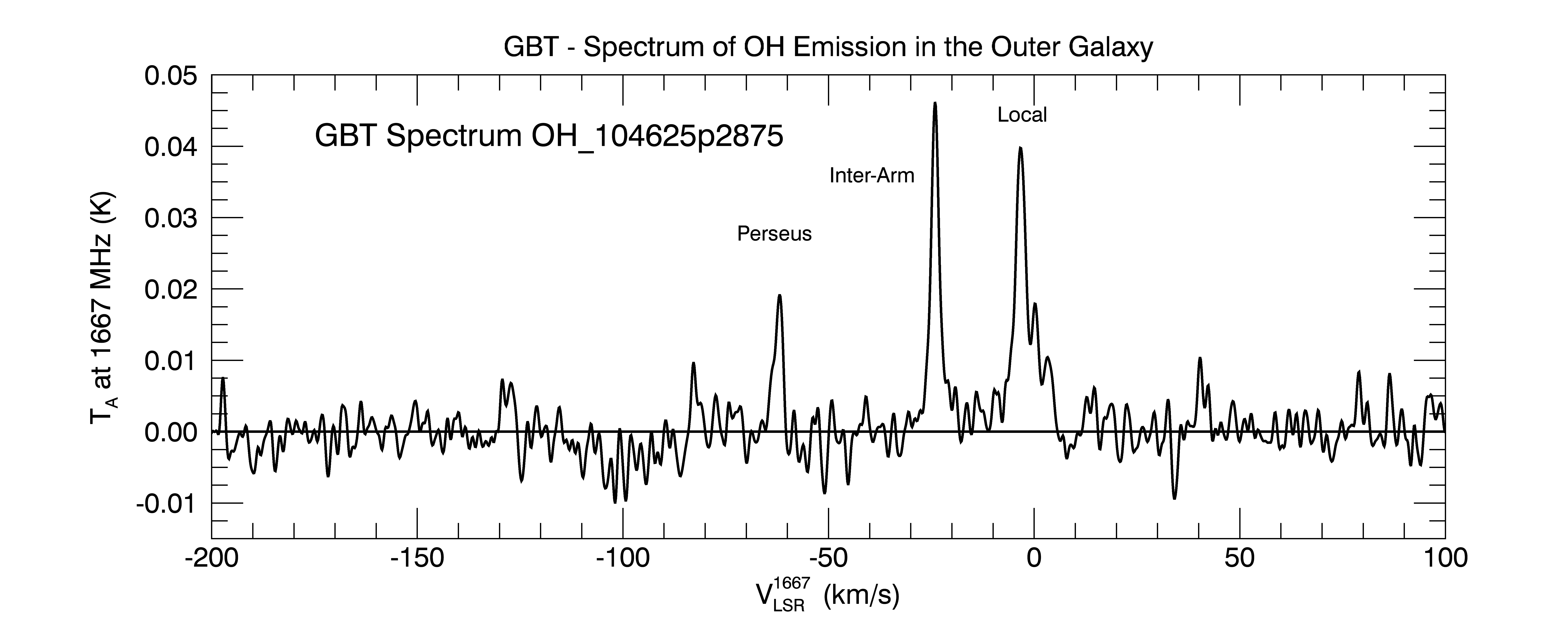}
\includegraphics[width=\textwidth]{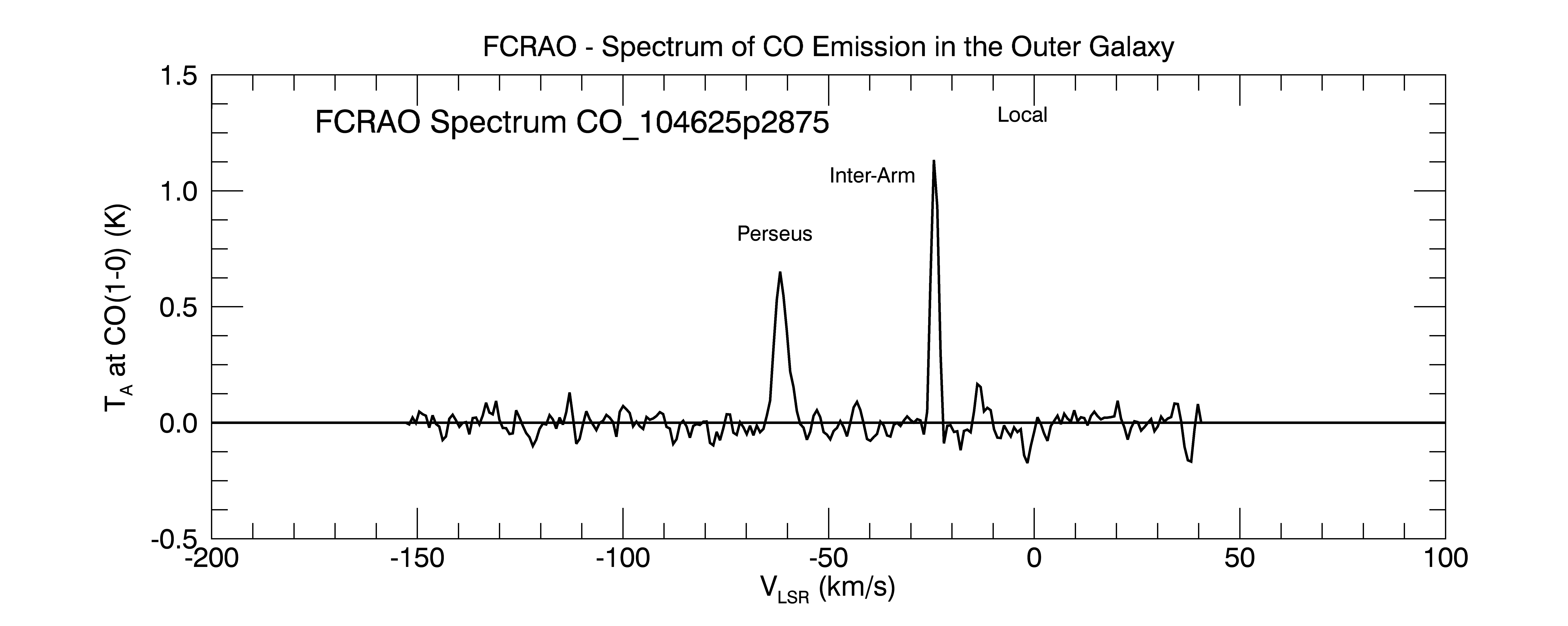}
\caption{Above: Example of a 1667 MHz OH spectrum from the One Square Degree survey. This spectrum was taken at $l = 104\fdg625, b = +2\fdg875$ with 2 hr of exposure time on the GBT. We consider any feature detected between -75 and -50 km s$^{-1}$ to be associated with the Perseus Arm. In this spectrum, the OH feature associated with Perseus Arm is located near -65 km s$^{-1}$. Smaller OH features, possibly spurious, can be observed sometimes at -80 km s$^{-1}$ and at -120 km s$^{-1}$, corresponding to the Outer Arm and Outer Scutum-Crux Arm respectively. The RMS on this spectrum as calculated by \textit{GTBTIDL} between -190 and -130 is 2.3 mK with a velocity resolution of 1 \kmps. Below: The corresponding example FCRAO CO spectrum in the same direction. Profile integrals are calculated over the same velocity range in both the CO and OH spectrum in the vicinity of the Perseus Arm. While this example does not demonstrate the CO-dark gas at the Perseus Arm velocity, it clearly does at local velocities. However, as shown in Fig. \ref{fig:heatMap}, a large portion of this survey is CO-dark at the velocity range of the Perseus Arm. The velocity resolution of the CO spectra is 0.8 \kmps, and the rms ranges between 60 mK to 100 mK.}
\label{fig:exampleSpectrum}
\end{figure*}

\subsection{Archival CO Data}

The \twCO(1-0) observations were made between 1995 and 1998 using the 14m telescope of the FCRAO \citep{Heyer1998TheGalaxy}. This FCRAO ``Outer Galaxy Survey'' observed a 336 square degree region of the second quadrant of the Galaxy sampled at $50\arcsec$ intervals with a FWHM beamwidth resolution of $45\arcsec$. For the purposes of our direct comparison, this survey was first smoothed to the GBT resolution. The final CO spectra have a sensitivity of 60 to 100 mK rms, and a velocity resolution of 0.8 \kmps.

\subsection{Data Reduction}

The GBT OH data were reduced using the \textit{GBTIDL} software \citep{Garwood2006GBTIDL:Data}. The quality of this data is generally very high, and each polarization of each 10 minute scan was reviewed for the presence of radio frequency interference and problems from instrumental effects. It was quickly evident that the spectra would be heavily influenced by the baseline ripple discovered during commissioning of the GBT (Fisher, J. R., Norrod, R.D., and Balser, D.S. 2003, Electronic Division Internal Report No. 312). The dominant feature is a broad ripple with an approximate periodicity of 9 MHz. Fisher et al. deduce that this ripple is caused by multipath reflections in which a part of the system noise enters the receiver system directly and a part is returned from the subreflector. Typical values of the amplitude are 0.4 K in Y and 0.1 K in X, the orthogonal polarized channel. Because the periodicity of this ripple is much larger than the frequency range expected for the OH signal, the ripple can in general be satisfactorily removed by fitting and subtracting a polynomial of low order to the baseline.

The residual spectrum shows an additional ripple feature that is more problematic. This ripple is comprised of several components having frequencies in a range between 1.3 and 1.8 MHz according to Fisher et al. The ripple is stronger in the Y polarization, is variable with a dependence on the configuration of the main reflector and the subreflector, and is presumed to also arise in multipath reflections from the circumferential gaps between the surface panels. The amplitude of the ripple is larger in the Y linear polarization whose E-vector is parallel to the plane of symmetry of the telescope. Typical worst case amplitudes are about $\pm 1.5$ mK in the Y polarization. The placement of the pattern in the bandpass is a strong function of the position of the subreflector. An attempt was made to reduce the amplitude of the ripple by observing the same position with the subreflector set at focus positions differing by $\lambda/8$. Some improvement was noted, but the ripple was never completely canceled out.

All of the data for a given position were assembled and reviewed for quality. Where necessary, data contaminated by (infrequent) interference or equipment instability were edited out. The data for each polarization were averaged, and the 9 MHz ripple was removed by fitting a baseline polynomial of order 5 to the spectrum in the frequency range 1661.4 MHz to 1671.4 MHz. Each spectrum was then used to compare the 1667 and 1665 MHz lines to judge if the emission is in LTE line ratio. The relative intensities of the four lines are 1:5:9:1 for the 1612-, 1665-, 1667-, and 1720-MHz lines respectively. Instances where the ratio differed from the LTE value proved to be rare. The comparison of the two spectra also provided a useful validation of faint spectral features.

The next step was to average the two polarizations in order to improve S/N, and to smooth the data by a Gaussian function to an effective spectral resolution of 1.0 \kmps. For these spectra the rms of an individual point is 2.2 mK, but there is variation from spectrum to spectrum because of the remaining uncertainty in the determination of the baseline. Including baseline uncertainties, the resultant RMS uncertainty in the OH spectra is on the order of 3mK in $T_{mb}$ units. The final spectrum was shifted to the rest frame velocity of the 1665 or 1667 MHz line for further detailed parameterization. An example spectrum from the OSD is shown in Fig. \ref{fig:exampleSpectrum}, the OH emission from the local gas, inter-arm, and Perseus spiral arm are observed at $V_{LSR}$ $\sim$ 0, -20 and -65 km s$^{-1}$ respectively.

The final step was to make the best estimate of OH emission from the Perseus Arm Region. We assumed that emission from the Perseus Arm would be limited to a spectral region of width 25 km/s centered on the expected velocity. For the expected velocity we were guided by the profile of the neutral hydrogen as mapped in the DRAO "LRDS" survey, in the vicinity of -75 to -50 \kmps\ \citep{Higgs2000ThePlane, Higgs2005TheII.}

To remove the residual effects of the 1.6 MHz ripple we defined two regions proximate to the OH window, each of width 10 km/s, and fitted a linear baseline to these regions. At the same time the CO emission from the region is computed from the archival data from the FCRAO \citep{Heyer1998TheGalaxy} over the same velocity range. The resulting parameters have been collected into a table, a fragment of which is shown in Table \ref{table:data}. The complete version of Table \ref{table:data} is provided in the electronic version of this paper.

The finished products from the \twCO\ data sometimes suffer from a long-wavelength ripple in the spectra, which has been removed by a similar polynomial fit. However, negative values are sometimes recovered owing to the variation in spectra where there is no CO signal (as is the case in most of the sightlines in the area studied here), see e.g. Figure 1 in \cite{Heyer1998TheGalaxy}, where there is a demonstrative negative feature, a probable artifact of this baseline ripple, at roughly the Perseus Arm velocity range.

\begin{figure*}[!ht]
\centering
\includegraphics[width=0.8\textwidth]{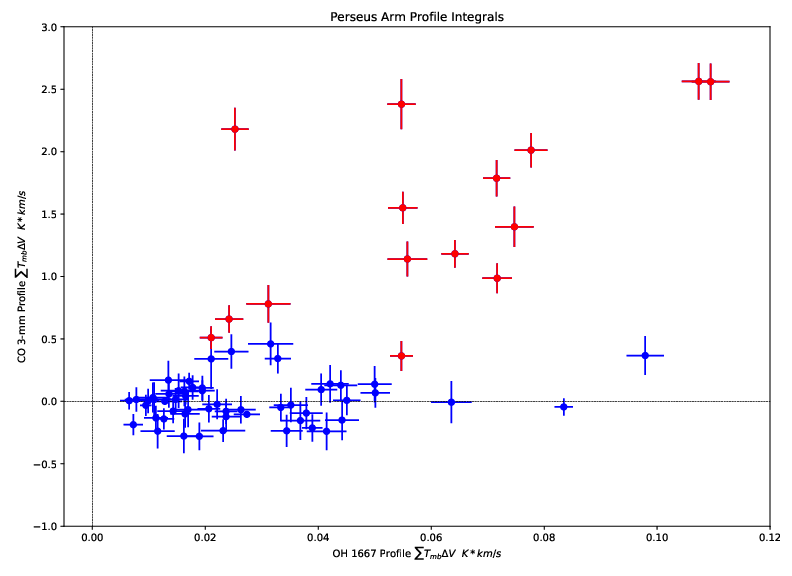}
\caption{The OH 1667 MHz emission line strength and the \twCO(1-0) line strength integrated over the same velocity range near $V_{LSR}\sim -65$ \kmps, corresponding the Perseus Arm. The pointings with detectable (above 3 $\sigma$) CO and OH are marked in red. Pointings only with detectable OH are marked in blue. There are 9 pointings with no detections in either OH or CO that are not shown on the plot. We note that there are no pointings that have a CO detections and an OH non-detection.}
\label{fig:profileIntegrals}
\end{figure*}

\subsection{Distance to the Perseus Arm}\label{distancetoarm}

The physical size of the Perseus Arm structures that we are observing are known from precise VLBI distance measurements  \citep{Reid2009TRIGONOMETRICMOTIONS, Reid2014TRIGONOMETRICWAY, Choi2014TRIGONOMETRICARM}. In the direction of $l = 105\fdg00, b = +2\fdg50$ we use the Bayesian Distance Calculator\footnote{http://bessel.vlbi-astrometry.org/bayesian} from The Bar and Spiral Structure Legacy Survey (BeSSeL) \citep{Reid2016ASources} to estimate the distance to our observed Perseus Arm sources. We note that the galactic rotation model utilized by the BeSSeL project does not incorporate the so-called ``rolling motions'' of the Perseus Arm \cite[]{Yuan1973TheArms, Foster2010StructurePicture}, and that the Bayesian Distance Calculator treats the Perseus Arm $V_{LSR}$ as constant with galactic latitude. While many of our observed features at high latitudes contain LSR radial velocities inconsistent with the Perseus Arm $V_{LSR}$ range provided by the BeSSeL model, we believe that rolling motions account for this variation in $V_{LSR}$ and that our observed features do indeed fall within the Perseus Arm. We therefore use the parallax distances from the calculator as opposed to the kinematic distance PDF; we adopt a distance estimate of 3.2 kpc to the Perseus Arm. As noted in the previous section, any feature we observe in the vicinity of -75 to -50 \kmps\ we associate with the Perseus Arm.

\subsection{Determining Line Strengths}

The line strengths of all features corresponding to the Perseus Arm velocity interval ($V_{LSR} \sim -65$ \kmps, see Fig. \ref{fig:exampleSpectrum}) of the 81 1667 MHz OH spectra in the survey were calculated. The line profile strengths of the CO and OH lines are plotted against each other in Fig. \ref{fig:profileIntegrals}. All 81 Perseus features in the OH spectra were manually identified and the velocity range's integration limits were chosen from visual inspection of the OH spectra. The corresponding line strengths from the CO spectra were obtained over the same velocity range. The expression used to calculate the line strengths is:

\begin{equation}
S = \sum T_{mb} \times \Delta V,
\end{equation}

\noindent in units of K \kmps, where $T_{mb}$ is in units of main-beam brightness temperature, and $\Delta V$ is the channel spacing of the data. Note that, for the GBT, $T_a$ and $T_{mb}$ are the same to within 5\%. The summation is done numerically over the channels containing measurable signal for each Perseus Arm feature in the 81 OH spectra. After this process, the \twCO(1-0) spectrum is extracted from a spatially-smoothed version of the FCRAO data cube and integrated over the same velocity range. This process was repeated for all 81 Perseus Arm features (and lack of features around $V_{LSR} \sim -65$ \kmps) until all of the data in the OSD was processed. The error analysis is the same procedure as outlined in \cite{Allen2015The+1deg} section 4.2. We also computed the scaled difference between the 1665 and 1667 MHz lines with each sightline and find that most emission is the LTE ratio of 5:9, as is expected if these lines are excited chiefly by collisions. 

\begin{figure*}[ht!] 
\vspace{0.2in}
\begin{center}
\includegraphics[width=0.4955\textwidth, angle=0]{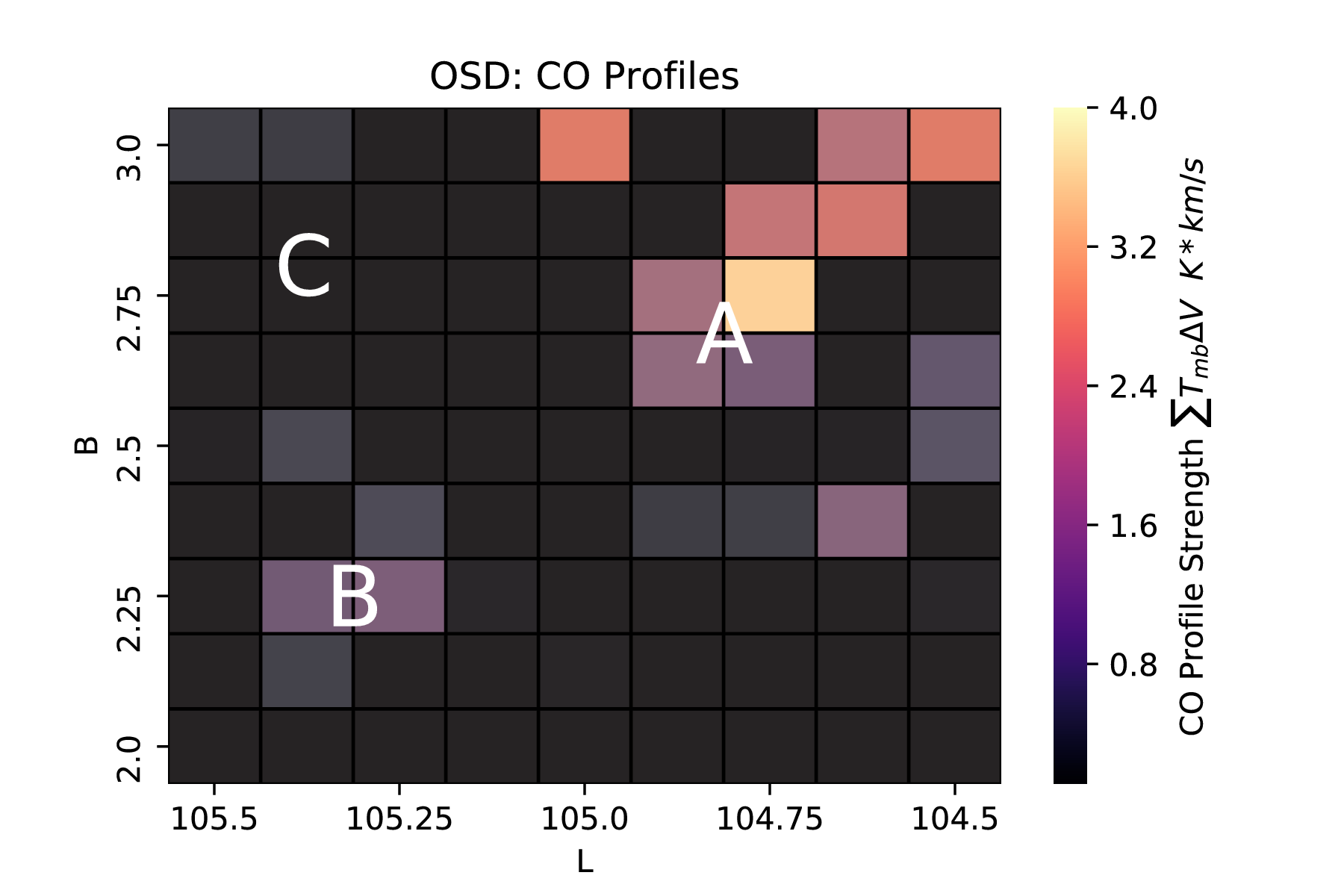} \hspace{0in} \includegraphics[width=0.4955\textwidth, angle=0]{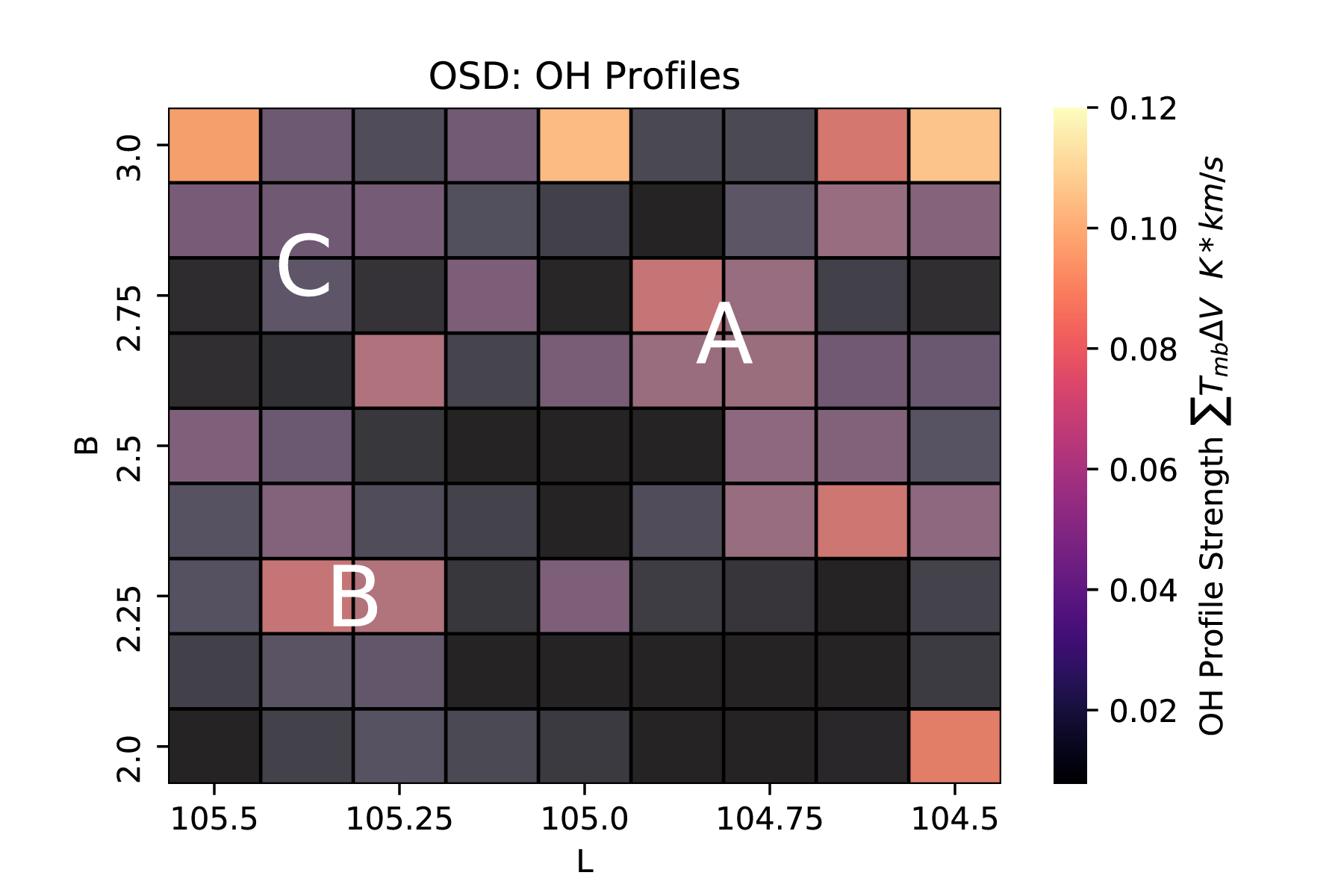}
\caption{The survey area in both molecular tracers. Left: The profile integral strengths for the CO line in the Perseus Arm velocity range ($V_{LSR} \sim -65$ \kmps). Right: The profile integral strengths for the OH 1667 MHz line over the same velocity range. The squares in these ``heatmaps'' are 7 pc on a side assuming they are at the distance of the Perseus Arm ($\sim$ 3.2 kpc). Extended OH emission is detected outside and surrounding the standard 'CO-bright' molecular gas, to distances on the order of 10-20 pc. The detections displayed here are in excess of $3\sigma_{rms}$ for both the CO and OH data; locations that are below that detection threshold are set to zero. The letters correspond to interesting features that we highlight, by eye, in Section 4.2.}
\label{fig:heatMap}
\end{center}
\end{figure*}

\subsection{Estimating the Continuum and Excitation Temperatures}

In order to fit our OH observations and the corresponding CO data to models of diffuse clouds in the ISM,  we need estimates of the column density of OH based on our line strength data. OH column densities are calculated from OH 1667 MHz emission lines using the following equation \cite[see, e.g.][]{Liszt1996GalacticSources}:
\begin{equation}
    N({\rm OH}) = C\frac{T_{ex}}{T_{ex} - T_C}\int{T_b(\nu) d\nu}\textbf,
\label{eq:emission}
\end{equation}
which in turn means that values for the continuum temperature in the background of the OH and of the excitation temperature of the 1667 MHz line are both needed as inputs. The coefficient $C = 2.3 x 10^{14}$ for the 1667 MHz line yields N(OH) in terms of cm$^{-2}$. However, determining the exact value of the continuum temperature is difficult, as observations have been made at 408 MHz, 1420 MHz, and 2695 MHz, but not at 1667 MHz; moreover, there are different components to the continuum, each of which have different spectral indices, making interpolation complicated. In order to estimate $T_C$ at 1667 MHz, we use the "diffuse component" of the continuum at 1420 MHz as reported by \citet{Reich1997The240}, and at 2695 MHz as reported by \citet{Furst1969A240DEG}, and we estimate the diffuse component from \cite{Haslam1969AMaps} as reported in \cite{Taylor2003TheSurvey}. We then made a quadratic interpolation on a logarithmic plot among the continuum temperature at these three frequencies to estimate the diffuse component of the continuum at 1667 MHz. Uncertainty still remains, however, since we do not know how much of the Galactic background portion of the continuum exists in the foreground versus the background of the OH. As such, we choose two possible values representative of the range of plausible $T_C$ values of 4.0 K and 5.0 K for this work. The value $T_C$ = 5.0 K assumes that the continuum is mainly in the background, and $T_C$ = 4.0 K assumes that half of the continuum is in the background, using \citet[]{Xu2016TheWay} Figure 2 as a rough basis.

The excitation temperature is another unknown quantity, but reasonable estimates are possible. Given that at the coordinate $104.75^{\circ}, 2.75^{\circ}$ CO is detected while the corresponding OH signal is very faint, and also that the source component of the continuum contains a small elevated patch in this vicinity which has been catalogued as the probable HII region GB6 B2214+5950 \cite[]{Gregory1996TheSources}, it seems plausible that at this coordinate the higher value of $T_C$ is close to the value of $T_{ex}$. That would put $T_{ex}$ at approximately 1 K above the surrounding value of $T_C$. Although this is the most likely scenario, since a $T_{ex}$ value closer to $T_C$ would have a significant effect on the resulting column densities, we also try a value of $T_{ex}$ only 0.5 K above the surrounding $T_C$ in our analysis.

\section{Results}\label{results}

Here we present the main results of this survey, beginning with a scatter plot of OH and CO profile integrals to demonstrate the presence of extended OH emission outside of CO-bright clouds in Fig. \ref{fig:profileIntegrals}. In addition, we display a 9 $\times$ 9 color-coded \textit{heatmap} based on profile integral strengths, which shows spatially the ubiquitous OH emission in the Perseus Arm in contrast to the relatively compact CO emission in Fig. \ref{fig:heatMap}. Each square in the heatmap represents a GBT beam which, at the assumed distance of the Perseus Arm, corresponds to roughly 7 pc in spatial extent. In a following subsection we present and discuss theoretical models of diffuse clouds which allow us to explore the physical environments of this gas in Fig. \ref{fig:cloudModel}.

\subsection{Cloud Model Predictions} \label{cloudModelSection}

\begin{figure*}[ht!]
    \centering
    \includegraphics[width=.46\textwidth]{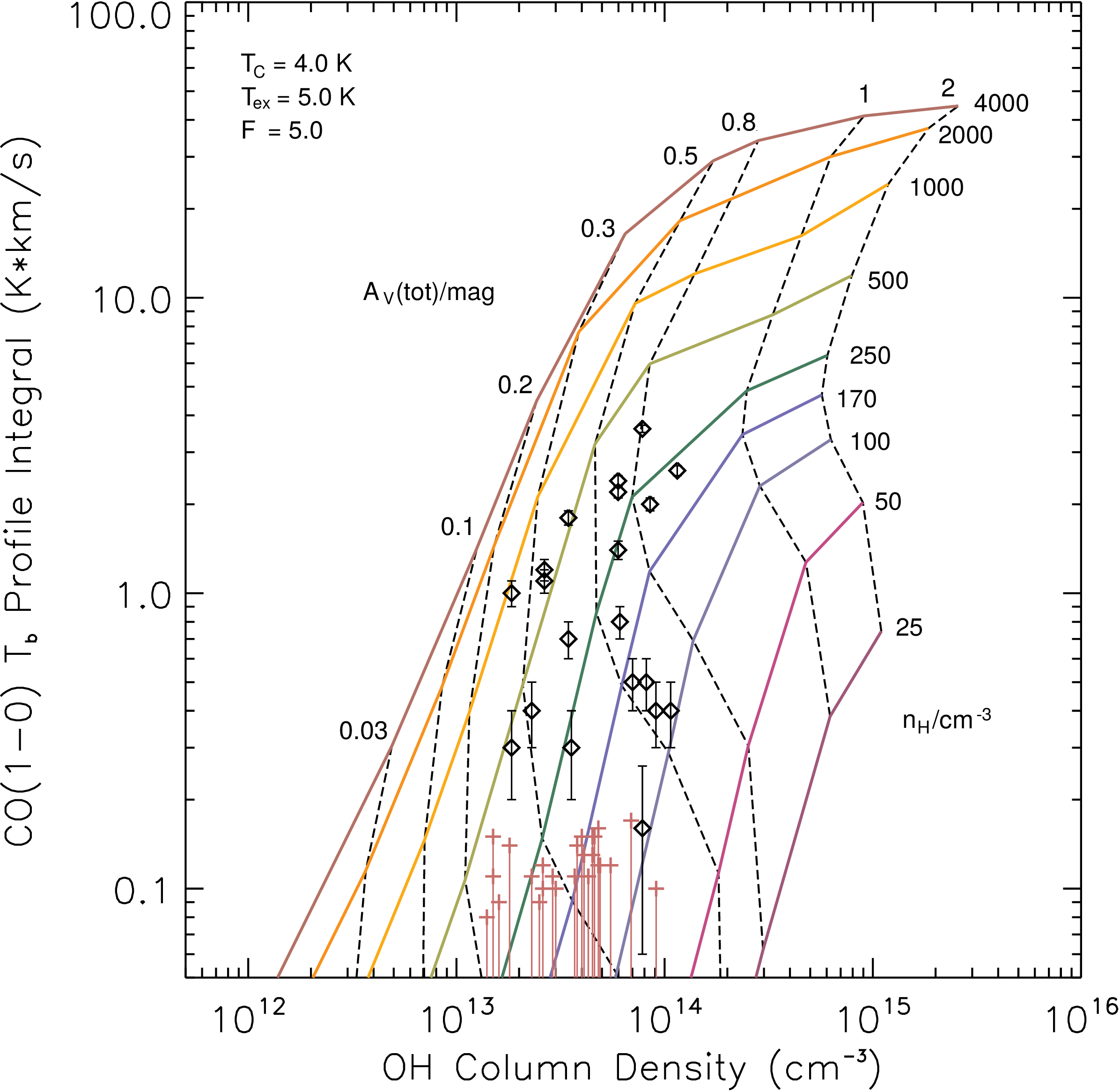}
    \includegraphics[width=.46\textwidth]{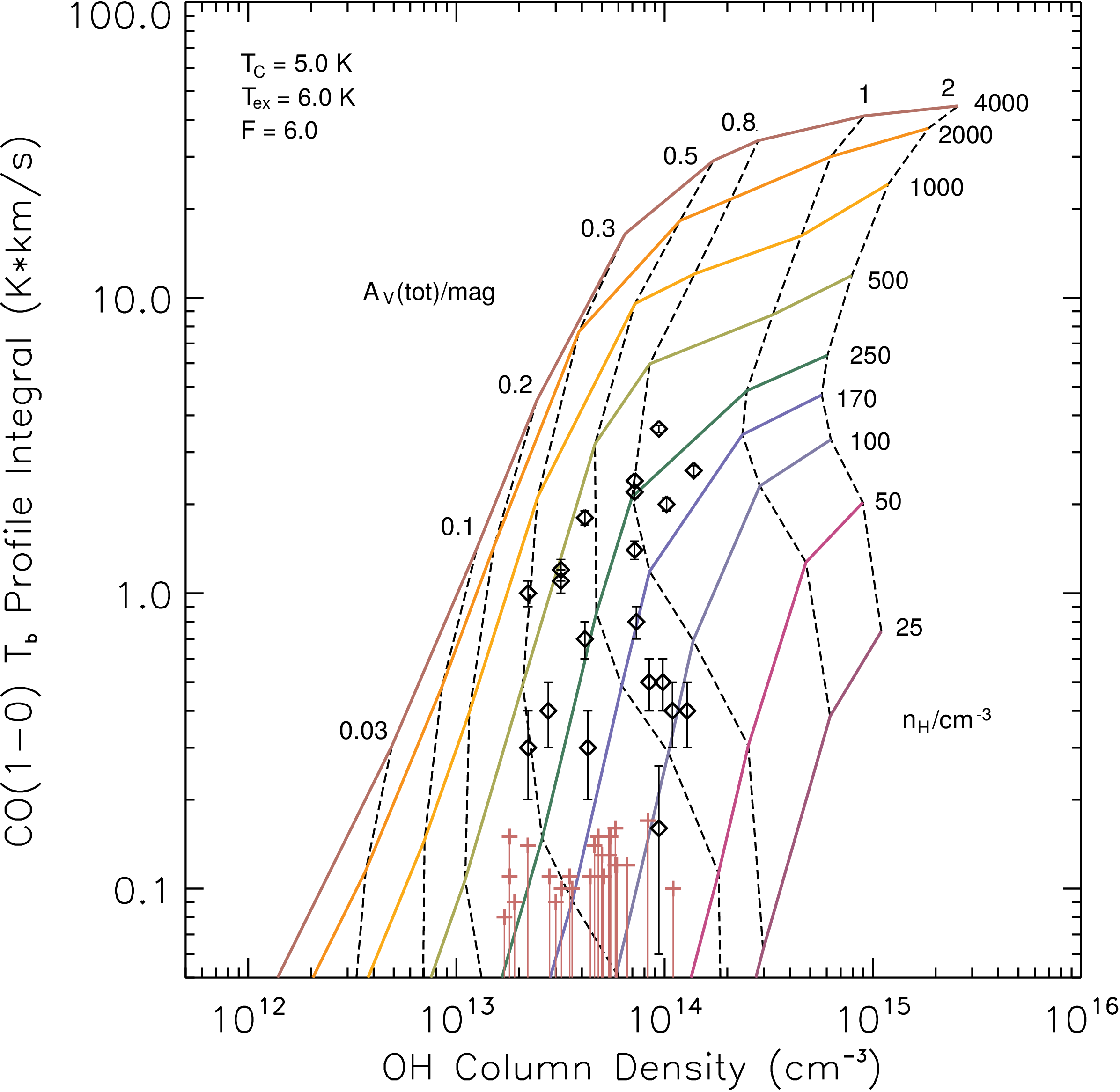}
    \includegraphics[width=.46\textwidth]{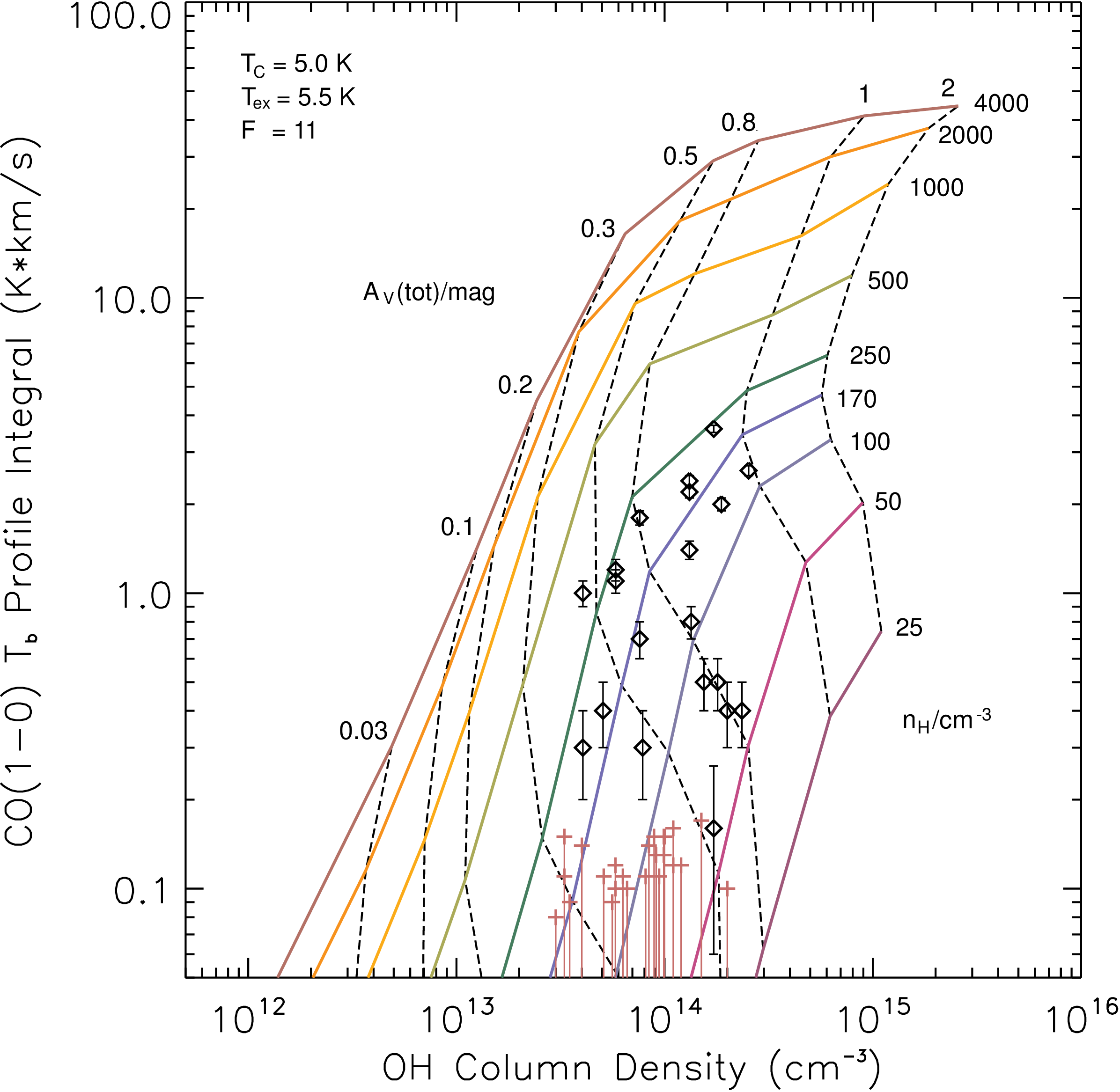}
    \caption{The predicted OH column densities and brightness temperatures of the CO(1-0) line as a function of the thickness of the cloud and the volume density of H nuclei ($n_{H}$ = $n_{HI}$ + 2 $n_{H_{2}}$). The colored contours indicate volume density of H nuclei, and the black dashed curves indicate thickness of the cloud in terms of visual extinction. OH Column densities calculated for each of our OH observations and the FCRAO Outer Galaxy CO data are plotted on the model curves. The data indicate that this region of the Perseus Arm exhibits low volume and low column densities and that the volume densities are typically lower for CO-dark gas than for CO-bright gas. The model was run with three sets of input values of $T_{C}$ and $T^{67}_{ex}$, and the factor F = $T^{67}_{ex}$/($T^{67}_{ex} - T_C$) representative of the plausible value range for this region. Error bars from the noise level on the OH observations are not shown but would be comparable in size to the size of the data markings.}
\label{fig:cloudModel}
\end{figure*}

We have compared the observed \twCO(1-0) and OH 1667 MHz profile integrals with the predictions of a set of diffuse cloud models. Here, we used the model described by \cite{Hollenbach2012TheIons}, with the modifications discussed by \cite{Neufeld2016TheClouds}, to obtain predicted CO profile integrals and OH column densities.  These were obtained as a function of the thickness of the cloud -- measured in magnitudes of visual extinction, $A_V({\rm tot})$ -- and the volume density of H nuclei, $n_{\rm H}$.  Results are shown in Fig. \ref{fig:cloudModel}, where colored contours representing fixed values of  $n_{\rm H}$ and black contours representing fixed values of $A_V({\rm tot})$ are plotted in the plane of observable quantities (i.e. the velocity-integrated brightness temperatures for the OH and CO transitions with the OH converted to column density.) All of the predictions shown here were obtained for an assumed cosmic-ray ionization rate of $2\times 10^{-16}\,\rm s^{-1}$ (primary ionizations per H atom) -- the mean Galactic value favored by \cite{Neufeld2017TheIons}--and an assumed interstellar ultraviolet radiation field equal to that given by \cite{Draine1978PhotoelectricGas}. 

The resulting values of gas volume density predicted by the cloud model for three representative values of the  factor F = $T^{67}_{ex}$/($T^{67}_{ex} - T_C$) are displayed in Table \ref{table:densities}. Note that for all three cases of input values of $T_C$ and $T^{67}_{ex}$, the predicted average volume density is greater for the CO-bright gas than it is for the CO-dark gas. As the CO-dark gas volume density predictions are upper limits, this result is further strengthened.

The diffuse cloud models suggest that the portion of the survey that is OH-bright, CO-dark gas is mainly molecular. According to the model, in the representative case where $A_V({\rm tot})$ = 0.3 mag and $n_{H}$ = 166 cm$^{-3}$, the molecular fraction of the gas is found to be 0.7. We also find that the predicted $A_V({\rm tot})$ is greater than $\sim 0.2, 0.25, 0.3$ mag for F = 5, 6, 11 respectively. This lower limit to $A_V({\rm tot})$ most likely results from the sensitivity of the observations and the required OH line strength to be detected above the noise.

\begin{table*}[ht!]
\begin{center}
\caption{H Nuclei Volume Density Results for Molecular Gas in the One Square Degree Survey. \label{table:densities}}
\begin{tabular}{c c c c c}
\hline
 $T_C$ (K) & $T^{67}_{ex}$ (K) & F & CO-Bright Mean Volume Density (cm$^{-3}$) & CO-Dark Mean Volume Density (cm$^{-3}$)\\
\hline
4.0 & 5.0 & 5.0 & $400 \pm 70$ & < $210 \pm 20$  \\
5.0 & 6.0 & 6.0 & 280 $\pm 40$  & < $200 \pm 20$\\
5.0 & 5.5  &  11  &  160 $\pm 15$ &  < $120 \pm 10$  \\
\end{tabular}
\end{center}
\end{table*}

\section{Discussion}\label{discussion}

\subsection{Comparing OH and CO Line Strengths}

Fig. \ref{fig:profileIntegrals} shows the 81 data points resulting from a profile analysis of the OH and CO features, over the same velocity ranges, in the Perseus Arm. OH emission appears widespread in this region and, if there is a CO signal detected, it is roughly correlated with the OH signal. It is apparent that a substantially larger region of the ISM contains molecular gas than is indicated by the \twCO\ line.

It is noteworthy to mention that the area of this survey (densely covering one square degree) and that of \citet[][sparsely covering four square degrees]{Allen2015The+1deg} are quite different, and yet they both statistically show very similar results: $40\%$ of the pointings in Paper I. showed 3$\sigma$ CO emission, while all of the pointings showed OH emission. In the present survey, $19\%$ of the analyzed pointings showed CO emission and $86\%$ showed OH emission. Both these results reveal large amounts of 'CO-dark' molecular gas in the ISM. The present survey is dense, showing structure on the scale of the GBT beam itself is 7 pc at the 3.2 kpc distance to the Perseus Arm in this direction, whereas the survey of \cite{Allen2015The+1deg} was not restricted to certain features in velocity space and was a sparse, blind OH survey that showed 'CO-Dark' gas structures on larger scales with large gaps between sight lines.

\subsection{Dark Gas Morphology on $0.125^{\circ}$ Intervals}\label{morphology}

The structure of dark gas is of interest for many reasons. Indirect observational evidence suggests that the diffuse molecular gas surrounds the 'CO-bright', molecular gas. Direct mapping of OH emission is extremely difficult due to sensitivity requirements but also shows observationally consistent results: the OH is extended beyond the CO-bright clouds \citep{Wannier1993WarmObservations, Allen2012Faint5circ, Allen2015The+1deg, Xu2016EvolutionTaurusb}. If one assumes that the OH molecule can trace all of the diffuse molecular gas, then the extent of the OH emission can tell us how extended this 'CO-faint' component of the molecular gas actually is. In this regard, we visually identified three interesting regions of note in Fig. \ref{fig:heatMap}: \

\begin{itemize}
\item Feature A: A CO-bright region centered on l = 104.7, b = 2.75, on the order of 12x12 pc, surrounded by extended OH emission;
\item Feature B: A CO-bright region centered on l = 105.25, b = 2.25. A dimmer and smaller structure than feature A, surrounded by extended OH emission, and;
\item Feature C: A CO-dark region around l = 105.25, b = 2.875 which  shows no CO emission but significant amounts of OH emission. This latter feature extends beyond feature A and B by at least 10-12 pc. 
\end{itemize}

In the case of Feature C, we note that there is clearly a CO-dark region with corresponding OH emission. Note also that our survey is also not fully sampled, that is: the intervals are beamwidths and not half-beamwidths. Finally, while the structure of the CO-bright regions are compact clouds which appear to be localized in our survey area, the structure of the OH emission is more diffuse and we cannot reliably conclude from our limited survey just how extended the dark gas actually is in relation to the CO-bright regions.  \cite{Grenier2005UnveilingNeighborhood} have described the `dark gas' as being `similar in extent to that of atomic HI; it appears to surround all CO-bright regions and bridges the dense cores to the more extended atomic distribution'. Our results show that OH emission is morphologically and quantitatively consistent with this description of the dark gas.

\subsection{The Physical Environment of the Dark Molecular Gas}

The physical state of the dark molecular gas is inferred to be diffuse clouds of \Htwo, without the necessary volume density to collisionally excite the 3mm CO(J=1-0) line. The Planck satellite detected an overabundance of IR emission in the column density ranges corresponding to $A_{v}$ < 2 \citep{PlanckCollaboration2011PlanckGalaxyb}. \cite{Wolfire2010TheGas} theoretically studied the dark gas phenomenon at the surface of molecular cloud surfaces and concluded that in this density range $H_{2}$ self-shields while CO can be photo-dissociated. As noted before, the OH-bright, CO-dark gas maintains a high molecular fraction ($\sim$ 0.7) in the models and remains a good candidate tracer of the dark molecular gas. However, for lower column and volume density regimes than studied in this paper, this may not be the case and further study is warranted.

Fig. \ref{fig:cloudModel} shows the diffuse cloud model predictions for the CO emission line strength and OH column densities, along with our Perseus Arm observations overplotted (see \ref{cloudModelSection}). We observe a density effect that is responsible for the CO-faint gas at the current sensitivities of the observations. In the context of these models, variations in the gas density are the most likely explanation for why some positions with detectable OH emission show CO emission above our detection threshold whereas other such positions do not.  A density dependence arises because the CO J = 1 state is subthermally populated over much of the density range of interest, and thus the rate of CO emission per molecule is roughly proportional to the density.  The OH lambda-doubling transitions, by contrast, become thermally-populated at much lower densities, owing to their much smaller spontaneous radiative decay rates.  Thus, the OH emission rate per molecule is independent of density, and the CO/OH line ratio is an increasing function of density in the regime of relevance.  Given the line sensitivities achieved in these observations, this behavior results in OH detections that are unaccompanied by CO detections when the gas density is low.

The results presented in Table \ref{table:densities} for the mean volume density of the CO-bright and CO-dark gas imply that, for the current level of sensitivity discussed in this paper, molecular gas becomes CO-dark below a volume density of H nuclei of $\sim 100-200$ \pcmcub\ in the survey region, which corresponds to a gas volume density of $\sim 50 - 100$ \pcmcub\ if the gas is primarily molecular hydrogen. The result that molecular gas becomes CO-dark below $\sim 50 - 100$ \pcmcub\ is consistent with the fact that the critical density for the \twCO(1-0) line is 1000 times larger than that of OH 18 cm transitions  \cite[see e.g.\ ][Chap. 16.2.1 and Fig. 16.1]{Wilson2013ToolsEdition}. Since the OH emission lines are observed to be in the LTE ratio of 5:9, collisions are expected to be the dominant source of excitation in this environment. The measured fraction of CO-dark molecular gas will depend on the detection threshold and velocity resolution of the comparison CO data set used \citep{Donate2017SensitiveTwo,Li2018WhereDMG}.

\section{Conclusions}\label{conclusions}

The results presented in this paper demonstrate that a substantial fraction of molecular gas is invisible in \twCO(1-0) emission at the current sensitivity level of the FCRAO survey, and this component of the molecular ISM can be effectively traced by sensitive 18-cm OH observations. 

\begin{itemize}
\item We presented a dense OH survey in the outer Galaxy with particular attention to the Perseus Arm molecular cloud velocity range ($V_{LSR}\sim -65$ \kmps). We showed that OH emission is widespread, which reaffirms previous observational results (see \cite{Allen2012Faint5circ, Allen2013ERRATUM:97, Allen2015The+1deg}), and can be spatially decoupled from CO emission.
\item We presented diffuse cloud models to predict the physical conditions of the molecular clouds studied in this paper. We demonstrated that the data, with reasonable assumptions for $T^{67}_{ex}$ and $T_C$, prefer theoretical densities that are well below the optically-thin critical density approximation for the \twCO(1-0) line. This offers a simple physical explanation for the simultaneous lack of observed CO emission, and yet the presence of extended OH emission in this region. The reason for this is because in the radio regime of low density molecular gas spectral line emission, stimulated emission is the dominant radiation mechanism \citep{Wilson2013ToolsEdition}.
\item The 18-cm OH lines appear to trace the behavior of the dark molecular gas, the CO-dark component of the molecular ISM that has been revealed from analysis of Planck data \citep{PlanckCollaboration2011PlanckGalaxyb}, gamma-ray \citep{Grenier2005UnveilingNeighborhood, Abdo2010FermiGalaxy, Ackermann2011TheTelescope}, and [CII] data \citep{Velusamy2010CIIClouds,Glover2016CO-darkGas}.
\end{itemize}

We also showed how CO and OH data, examined jointly with simulations of the diffuse ISM, can reveal the physical conditions of the dark gas. The large-scale distribution of 18-cm OH emission in the galaxy could shed light on the features of the galactic molecular ISM not revealed by conventional CO mapping observations. Future observations by the 500m FAST telescope \citep{Li2016TheProject}, and the SKA \citep{Dickey2012GASKAPSurvey} may help to reveal a portion of this dark molecular ISM content through OH absorption studies; however, the low excitation temperature of the 18-cm OH lines \citep{Engelke2018OHW5} remains an impediment to the straightforward interpretation of OH observations in the inner Galaxy in the presence of substantial ambient nonthermal continuum emission. An all-sky survey for OH emission in the outer Galaxy would not be hindered by the much weaker continuum of the galactic background there; however, an effective OH emission survey may be presently out of reach of current instruments owing to the long integration times required to obtain the necessary sensitivity. Nevertheless, such a survey could be  critical for an understanding the true extent of diffuse molecular gas in the outer Galaxy.

\acknowledgments

We are grateful to the staff at the Green Bank Observatory for their advice and assistance with the operation of the GBT, in particular Karen O'Neil, Jay Lockman, Toney Minter, Ron Maddalena, and Amanda Kepley, and for the development and support of the GBTIDL data analysis software, especially Jim Braatz and Bob Garwood. We would like thank Joanne Dawson, Claire Murray, Anita Petzler, Katie Jameson and Bruce Draine for helpful and insightful discussions, and Mark Heyer for providing an updated digital copy of the FCRAO data. We also want to thank the anonymous referee for constructive comments that helped improve this manuscript. M. P. B. is supported by a National Science Foundation Graduate Research Fellowship under grant No. 1746891. The work of D. A. N. and M. G. W.  was supported by grant No. 120364 from NASA's Astrophysical Data Analysis Program (ADAP). This research has been supported by the Director's  Research Fund at STScI.

\software{Astropy (The Astropy Collaboration 2013, 2018), GBTIDL (Garwood et al. 2006)}

\bibliographystyle{aasjournal}
\bibliography{Mendeley}

\begin{deluxetable}{cccccccccc}

\rotate



\tablecaption{OH and CO Profile Integrals in the current ``One-Square-Degree'' (OSD) Survey \label{table:data}}

\tablenum{2}

\tablehead{\colhead{GLON} & \colhead{GLAT} & \colhead{OH Int} & \colhead{OH Int Err} & \colhead{OH Med. Vel.} & \colhead{OH Med. Vel. Err} & \colhead{CO Int} & \colhead{CO Int Err} & \colhead{CO Med. Vel.} & \colhead{CO Med. Vel. Err} \\ 
\colhead{(deg)} & \colhead{(deg)} & \colhead{(K*km/s)} & \colhead{(K*km/s)} & \colhead{(km/s)} & \colhead{(km/s)} & \colhead{(K*km/s)} & \colhead{(K*km/s)} & \colhead{(km/s)} & \colhead{(km/s)} } 

\startdata
104.5 & 3 & 0.104 & 0.0033 & -64.019 & 0.092 & 2.56 & 0.146 & -64.21 & 0.4 \\
104.625 & 3 & 0.0738 & 0.0029 & -63.49 & 0.185 & 2.012 & 0.137 & -64.22 & 0.406 \\
104.75 & 3 & 0.0185 & 0.0022 & -61.11 & 0.185 & 0.085 & 0.084 & -61.78 & 1.21 \\
104.875 & 3 & 0.018 & 0.0025 & -61.29 & 0.64 & -0.28 & 0.11 & -61.37 & 4.06 \\
105 & 3 & 0.102 & 0.003 & -64.0192 & 0.09 & 2.563 & 0.146 & -64.2199 & 0.406 \\
105.125 & 3 & 0.0334 & 0.003 & -63.162 & 0.55 & -0.0294 & 0.137 & -63.81 & 5.68 \\
105.25 & 3 & 0.0196 & 0.002 & -61.67 & 0.277 & -0.0596 & 0.109 & -60.9 & 3.25 \\
105.375 & 3 & 0.0312 & 0.0023 & -66.137 & 0.277 & 0.3428 & 0.119 & -67.47 & 0.812 \\
105.5 & 3 & 0.093 & 0.0033 & -65.59 & 0.185 & 0.367 & 0.1547 & -55.65 & 0.812 \\
104.5 & 2.875 & 0.0428 & 0.0024 & -62.998 & 0.185 & 0.00791 & 0.119 & -63.41 & 2.03 \\
104.625 & 2.875 & 0.052 & 0.0025 & -62.26 & 0.185 & 1.74 & 0.109 & -61.78 & 0.406 \\
104.75 & 2.875 & 0.024 & 0.0024 & -62.47 & 0.277 & 0.377 & 0.109 & -60.96 & 0.4 \\
104.875 & 2.875 & 0.0069 & 0.0017 & -62.66 & 0.55 & -0.187 & 0.084 & -62.59 & 1.62 \\
105 & 2.875 & 0.0156 & 0.00259 & -61.082 & 0.55 & -0.101 & 0.109 & -31.29 & 62.59 \\
105.125 & 2.875 & 0.021 & 0.0026 & -61.86 & 0.46 & -0.0239 & 0.119 & -60.56 & 4.06 \\
105.25 & 2.875 & 0.035 & 0.0035 & -62.43 & 0.277 & -0.154 & 0.154 & -63.81 & 7.31 \\
105.375 & 2.875 & 0.0317 & 0.002 & -63.05 & 0.185 & -0.049 & 0.109 & -32.1 & 64.21 \\
105.5 & 2.875 & 0.036 & 0.0029 & -51.9 & 0.37 & -0.094 & 0.128 & -52.0314 & 4.87 \\
... & ... & ... & ... & ... & ... & ... & ... & ... & ... \\
\enddata


\tablecomments{A fragment of a table of results from the data reduction described in section 2 is shown here. The full version of this table will be available in the electronic version of this journal.}

\end{deluxetable}

\end{document}